\documentclass[prb,aps,twocolumn,superscriptaddress]{revtex4}
\usepackage{graphicx,amsmath,bm,xcolor}
\usepackage{dcolumn}

\begin{document}

\title{Hidden Bose-Einstein Singularities in Correlated Electron Systems:\\
 II. Pseudogap Phase in the Weakly Attractive Hubbard Model
}

\author{Takafumi Kita}
\affiliation{Department of Physics, Hokkaido University, Sapporo 060-0810, Japan}
\date{\today}

\begin{abstract}
The hidden Bose--Einstein singularities of correlated electron systems, 
whose possible existence has been pointed out in a previous paper
based on quantum field theory of ordered phases [T. Kita, J. Phys. Soc. Jpn. {\bf 93}, 124704 (2024)], 
are studied in more detail in terms of the attractive Hubbard model, for which 
the mean-field theory predicts that spin-singlet superconductivity is realized at low enough temperatures for any band structure
and interaction strength.
It is shown that incorporating correlation effects should change the mean-field superconducting solution
substantially and qualitatively even in the weak coupling, implying that the system lies in the strong-coupling region perturbatively.
The hidden singularity is found to be present
around the mean-field superconducting temperature $T_{{\rm c}0}$, below which
the standard self-consistent treatment by quantum field theory cannot be used due to divergences in the zero Matsubara frequency 
branch obeying Bose--Einstein statistics.
Our method to recover the applicability with a Lagrange multiplier
predicts that the singularity is a physical entity signaling the threshold of a pseudogap phase 
with a characteristic V-shape structure in the density of states near zero energy,
which lies above the superconducting phase and 
originates from the emerging one-particle-reducible structure in the self-energy.
\end{abstract}

\maketitle

\section{Introduction}

Self-consistency is an essential ingredient for describing any ordered phase that emerges spontaneously based 
on some nonlinear equation.
Among the most successful of those equations are the gap equation of the Bardeen-Cooper-Schrieffer theory\cite{BCS57} and its inhomogeneous extension called the Bogoliubov-de Gennes or Andreev equations.\cite{Bogoliubov59,deGennes66,Andreev64}
Indeed, these mean-field equations have been shown to reproduce properties of metallic superconductors 
excellently and quantitatively.\cite{AGD63,Parks69,Kita15}
Self-consistency is also a key feature of the Landau theory of Fermi liquids,\cite{Landau56,Landau57,BP91}
where interactions between particles are handled as an average molecular field within the mean-field framework.\cite{AGD63,Kita15,BP91,Leggett75,SR83}
Moreover, the two kinds of self-consistency have been shown to be able to describe basic properties of 
both the normal and superfluid phases of liquid $^3$He,\cite{Leggett75,SR83} even though the bare interaction between particles is 
quite strong there.
Hence, it is desirable to have a self-consistent framework that enables us to incorporate many-body correlations
systematically on top of the mean-field results so as to quantitatively describe normal and ordered phases with considerable correlations 
like high-$T_{\rm c}$ superconductors\cite{BM86,TS99,Yanase03,NPK05,Kontani13} on the same footing.

Such a formalism was developed by Luttinger and Ward,\cite{LW60} who succeeded in expressing the grand thermodynamic potential
$\Omega$ of the normal phase as a functional of Green's function $G$.
It can be regarded as resulting from a Legendre transformation 
of $\Omega$ from a non-local external potential $U_{\rm ext}$ to Green's function $G$,
hence obeying the stationarity condition $\delta \Omega/\delta G=0$ for $U_{\rm ext}=0$.\cite{DDM64,JL64}
The Luttinger--Ward (LW) functional is alternatively called quantum effective action
in relativistic field theory,\cite{CJT74,Weinberg96} which has been used for proving several fundamental theorems
such as the Fermi-surface sum rule for the normal state\cite{Luttinger60} and 
Goldstone's theorem on spontaneously broken continuous symmetries.\cite{GSW62,JL64}
Moreover, a part of the LW functional called {\it the $\Phi$ functional}
can be used as a generator of systematic approximations beyond the self-consistent Hartree-Fock theory, 
as already noted by Luttinger,\cite{Luttinger60}
which were shown by Kadanoff and Baym\cite{Baym61,Baym62} to have a remarkable property indispensable for 
describing nonequilibrium phenomena of satisfying various conservation laws automatically;
the original formulation on the imaginary-time Matsubara contour has been extended onto the real-time Keldysh contour\cite{Keldysh64,RS86,HJ98,Kita10}
with presenting  definite Feynman rules on it and 
also clarifying the hierarchical structure from the Schwinger-Dyson equation through the Boltzmann equation
down to the Navier--Stokes equations.\cite{Kita10}
These facts show how useful the LW formalism is, which has another unique (but not widely appreciated) value of being
able to describe ordered phases besides the normal phase through its self-consistency procedure inherent in the formalism.\cite{DDM64}
Noting this point, we have made efforts to extend the LW formalism so as to describe superconductivity with correlations concisely
as a natural extension of the BCS theory\cite{Kita96,Kita11} and also
Bose--Einstein condensation in such a way as to obey Goldstone's theorem.\cite{Kita09,Kita14,Kita21-1,Kita21-2}
It has thereby been shown\cite{Kita11}  that ordered phases of fermionic systems, such as superconductivity and ferromagnetism,
can be described in terms of Feynman diagrams without arrows on Green's function lines by using a single vertex with particle-hole and spin degrees of freedom; see Fig.\ \ref{fig2} below on this point.
This has resulted in a substantial reduction of the number of Feynman diagrams to be drawn and
also a natural inclusion of all the anomalous processes of superconductivity derivable from a given normal-state diagram.
Note also the advantage of the LW formalism over those starting from either 
the $\Phi$ functional, Green's functions, or response functions that we can calculate the thermodynamic potentials directly.

On the basis of this formalism for correlated superconductors,\cite{Kita11} 
it was shown previously\cite{Kita24} that there may exist a new kind of singularities
called {\it  hidden Bose--Einstein (BE) singularities}.
Specifically, they possibly lie around the mean-field transition temperature $T_{{\rm c}0}$ of every ordered phase 
in the zero Matsubara frequency branch of correlation contributions to the grand thermodynamic potential $\Omega$.
When the singularity is hit and passed down,
the standard self-consistent framework of quantum field theory can no longer yield any definite solution due to the divergence of the self-energy.
To overcome the difficulty, we have incorporated the physical condition based on the method of Lagrange multipliers 
that the argument of the logarithm in the correlation contribution to $\Omega$ cannot be negative in the zero Matsubara frequency branch.
The condition in turn is predicted to bring a novel structure in the self-energy called {\it one-particle reducible} (1PR),
which causes a change in the single-particle density of states without accompanying broken symmetries.
However, the numerical results were obtained only at an intermediate coupling strength
for a specific band structure of the quadratic dispersion with a large cutoff energy compared with the Fermi energy,
thereby leaving unclear how general the statement is.

With these backgrounds, the aims of the present paper are twofold. 
The first one is to clarify in more detail under what conditions
the hidden BE singularities emerge.
To this end, we focus on the attractive Hubbard model in three dimensions,
which has the definite advantage of yielding a spin-singlet superconducting solution within the mean-field analysis 
even in the weak-coupling limit for any band structure.
We will study how the mean-field superconducting solution is altered as we switch on correlation effects.
It will thereby be shown that reaching the mean-field transition temperature $T_{{\rm c}0}$ signals 
that the system enters the region where some non-perturbative treatment of the correlation effects is required,
however small $U$ is.
By reducing the temperature further we encounter the hidden BE singularity, below which the standard self-consistent treatment of 
quantum field theory cannot be applicable.
This fact implies that the hidden BE singularity is a general feature of the attractive Hubbard model.
The second purpose concerns our method developed for describing the system below the singular point
based on the method of Lagrange multipliers. 
We will obtain an analytic expression of Green's function with the resulting 1PR structure in the self-energy 
in terms of conventional Green's functions,
which will enable us to calculate the density of states more easily through the phase transition point of broken symmetries.

We adopt the units of $\hbar=k_{\rm B}=1$ throughout.

\section{Hidden Bose--Einstein Singularity}

\subsection{Model}

We will consider the attractive Hubbard model
\begin{align}
\hat{H}=\sum_{{\bf k}\sigma}\xi_{{\bf k}}\hat{c}_{{\bf k}\sigma}^\dagger\hat{c}_{{\bf k}\sigma}+U
\sum_{j}\hat{c}_{j\uparrow}^\dagger\hat{c}_{j\uparrow}\hat{c}_{j\downarrow}^\dagger\hat{c}_{j\downarrow}
\end{align}
with $U<0$. Here, ${\bf k}$ and $\sigma\!=\,\uparrow,\downarrow$ are the Bloch wave vector and spin quantum number, respectively, 
$\xi_{\bf k}$ is the single-particle energy in a finite band measured from the chemical potential $\mu$, $(\hat{c}_{{\bf k}\sigma}^\dagger,\hat{c}_{{\bf k}\sigma})$ are the creation-annihilation operators obeying the anticommutation relations, and $j=1,\cdots, {\cal N}_{\rm a}$ specifies the lattice site ${\bf R}_j$ with 
\begin{align*}
\hat{c}_{j\sigma}=\frac{1}{\sqrt{{\cal N}_{\rm a}}}\sum_{\bf k} \hat{c}_{{\bf k}\sigma}\,e^{i{\bf k}\cdot {\bf R}_j}.
\end{align*}

\subsection{Mean-field transition temperature}

The mean-field superconducting transition temperature $T_{{\rm c}0}$ of the spin-singlet pairing
is determined by the linearized self-consistency equation\cite{BCS57,Bogoliubov59,deGennes66,AGD63,Parks69,Kita15}
\begin{subequations}
\label{Tc0-eq}
\begin{align}
-\frac{U}{{\cal N}_{\rm a}}\sum_{\bf k} \frac{1}{2\xi_{\bf k}}\tanh\frac{\xi_{\bf k}}{2T_{{\rm c}0}} =1 .
\label{Tc0-eq1}
\end{align}
It can be written alternatively by using non-interacting Greens function $G_0(\vec{k})\equiv (i\varepsilon_n-\xi_{{\bf k}})^{-1}$ as
\begin{align}
-\frac{U}{{\cal N}_{\rm a}}T\sum_{\vec{k}} G_0(\vec{k})G_0(-\vec{k})\biggr|_{T=T_{{\rm c}0}}=1,
\label{Tc0-eq2}
\end{align}
where $\vec{k}\!\equiv \!({\bf k},i\varepsilon_n)$ with $\varepsilon_n\!\equiv\! (2n+1)\pi T$ ($n\!=\!0,\pm 1, \cdots$) denoting the fermion Matsubara frequency.
Indeed, performing the summation over $n$ in Eq.\ (\ref{Tc0-eq2})\cite{LW60} reproduces Eq.\ (\ref{Tc0-eq1}).
Moreover, we can express Eq.\ (\ref{Tc0-eq2}) concisely as
\begin{align}
-U\chi_{\rm pp}^0(\vec{0})\Bigr|_{T=T_{{\rm c}0}}=1
\label{Tc0-eq3}
\end{align}
\end{subequations}
in terms of the particle-particle bubble defined by
\begin{align}
\chi_{\rm pp}^0(\vec{q})\equiv \frac{T}{{\cal N}_{\rm a}}\sum_{\vec{k}}G_0(\vec{k}+\vec{q})G_0(-\vec{k}),
\label{chi_pp}
\end{align}
where $\vec{q}\equiv ({\bf q},i\omega_\ell)$ with
$\omega_\ell\!\equiv\! 2\ell\pi T$ ($\ell\!=\!0,\pm 1, \pm 2,\cdots$) denoting the boson Matsubara frequency.
One can show that
$\chi_{\rm pp}^0({\bf q},0)$ in the zero Matsubara frequency branch is real  through a change of variable $n\rightarrow -n$.

Thus, we have seen that the equation for $T_{{\rm c}0}$ can be expressed with respect to a
product of two normal-state Green's functions. 
This is a general feature of the mean-field equations for 
transition temperatures of ordered phases that are realized continuously,
as shown in Appendix\ref{AppA}.

\subsection{Non-negligible corrections of higher orders near $T_{{\rm c}0}$\label{subsec:PE}}

\begin{figure}[b]
\centering
\includegraphics[width=0.95\linewidth]{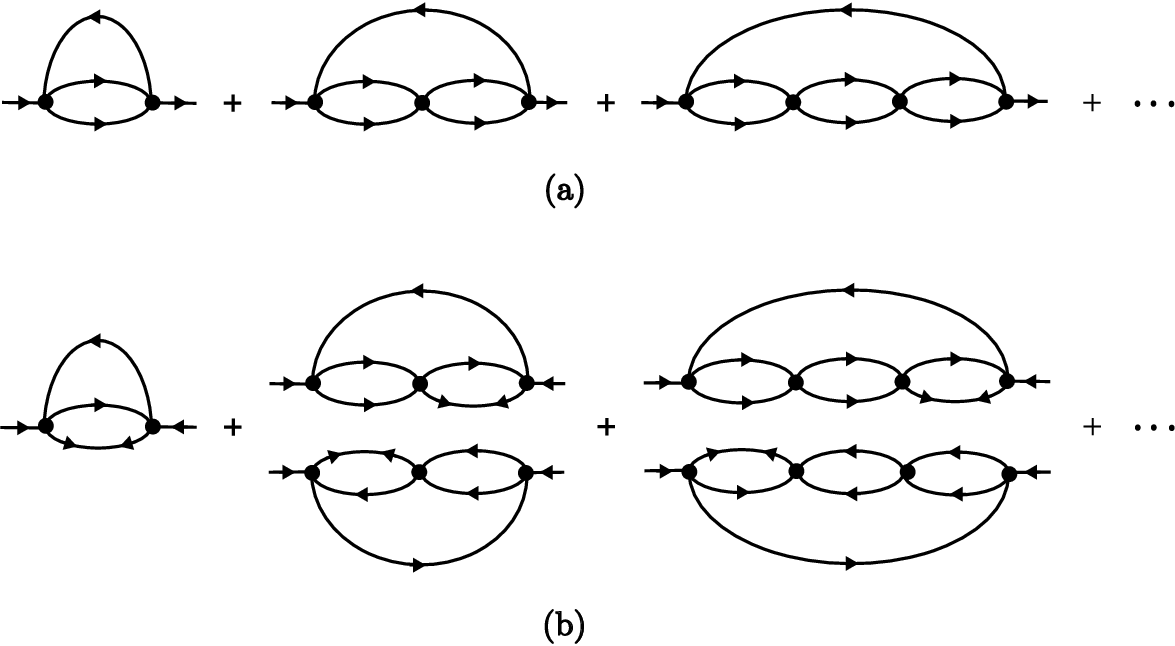}
\caption{\label{fig1}(a) Diagrammatic expression of the normal self-energy consisting of particle-particle bubbles; a filled circle denotes $U$.\cite{AGD63}
(b) Anomalous self-energy diagrams composed of particle-particle bubbles that are first order in $F$ 
and contribute to the $T_{\rm c}$ equation,
where Eq.\ (\ref{dF}) should be used in place of $F$.
}
\end{figure}

One may think that Eq.\ (\ref{Tc0-eq}), which is first order in $U$, is excellent quantitatively 
to estimate the superconducting transition temperature of the attractive Hubbard model in the weak coupling. 
However, Eq.\ (\ref{Tc0-eq3}) implies that there appears a series 
in the perturbative treatment of the self-energy which should be summed up to the infinite order.
Specifically, the normal self-energy $\Sigma$ contains the series given diagrammatically as Fig.\ \ref{fig1}(a),
which can be expressed analytically in terms of the particle-particle bubble of Eq.\ (\ref{chi_pp}) as
\begin{align}
\Sigma_{\rm pp}^0(\vec{k}) \equiv &\, U \frac{T}{{\cal N}_{\rm a}}\sum_{\vec{q}} \sum_{n=1}^\infty \bigl[-U\chi_{\rm pp}^0(\vec{q})\bigr]^n G_0(-\vec{k}+\vec{q})
\notag \\
=&\,U \frac{T}{{\cal N}_{\rm a}}\sum_{\vec{q}} \frac{-U\chi_{\rm pp}^0(\vec{q})}{1+U\chi_{\rm pp}^0(\vec{q})}
G_0(-\vec{k}+\vec{q}).
\label{Sigma_pp0^}
\end{align}
This contribution is divergent around $\vec{q}=\vec{0}$ for $T\gtrsim T_{{\rm c}0}$
according to Eq.\ (\ref{Tc0-eq3}) so that it cannot be cut at any finite order.
The fact also suggests that Green's function 
\begin{align}
G(\vec{k})=\left[i\varepsilon_n-\xi_{\bf k}-\Sigma(\vec{k})\right]^{-1}
\label{G(k)}
\end{align}
may deviate substantially from $G_0(\vec{k})=(i\varepsilon_n-\xi_{\bf k})^{-1}$ to make the estimation of 
$T_{{\rm c}0}$ by Eq.\ (\ref{Tc0-eq2}) inadequate quantitatively.

Moreover, the linearized self-consistency equation (\ref{Tc0-eq2}) should also be supplemented with the series of 
Fig.\ \ref{fig1}(b). It is thereby replaced by 
the eigenvalue problem that
\begin{align}
-\frac{T}{{\cal N}_{\rm a}}\sum_{\vec{k}'}U_{\vec{k}\vec{k}'}^{\rm s0}G_0(\vec{k}')G_0(-\vec{k}')\Delta(\vec{k}') =\epsilon \Delta(\vec{k})
\label{Tc0-eq4}
\end{align}
has an eigenvalue $\epsilon=1$ at the transition temperature, where $U_{\vec{k}\vec{k}'}^{\rm s0}$ is given analytically by
\begin{align}
U_{\vec{k}\vec{k}'}^{\rm s0}=&\,U-U^2\frac{T}{{\cal N}_{\rm a}}\sum_{\vec{q}}\left[\frac{1}{1+U\chi_{\rm pp}^0(\vec{q})}-\frac{1}{2}\right]
\notag \\
&\,\times
[G(\vec{k}+\vec{q})+G(-\vec{k}+\vec{q})]G(\vec{k}'+\vec{q}).
\label{U^s0}
\end{align}
Indeed, Eq.\ (\ref{Tc0-eq2}) is reproduced from Eq.\ (\ref{Tc0-eq4}) with $\epsilon=1$
by omitting the second term of Eq.\ (\ref{U^s0}).
However, this second term cannot be neglected
compared with the first one for $T\lesssim T_{{\rm c}0}$. 
This fact suggests that there may be no region in the attractive Hubbard model in which the mean-field treatment of superconductivity
is appropriate quantitatively, and conceivably, even qualitatively.

Note in this context that, when $-U\chi_{\rm pp}^0(\vec{0})\!>\!1$ is satisfied, Eqs.\ (\ref{Sigma_pp0^}) and (\ref{U^s0})
can no longer yield any physical solution due to divergences of 
the sums over ${\bf q}$ at $\omega_\ell=0$.
Indeed, the condition $-U\chi_{\rm pp}^0(\vec{0})>1$ implies that 
the Taylor series in the $\omega_\ell=0$ branch of Eq.\ (\ref{Sigma_pp0^}) 
does not converge over a finite region of ${\bf q}$ to make $\Sigma_{\rm pp}^0(\vec{k})$ indefinite;
the same statement holds for Eq.\ (\ref{U^s0}).

\subsection{Connection with previous studies}

The series of Eq.\ (\ref{Sigma_pp0^}) has generally been called {\it the $T$-matrix approximation};\cite{Baym61,Baym62}
both the bare series in terms of $G_0$ and its self-consistent version with $G$
have been used extensively in the literature to incorporate correlation effects of particle-particle bubbles in the normal self-energy.\cite{NSR85,Haussmann93,Haussmann94,HRCZ07,Yanase03,CSTL05,KWO12,SPRSU18,BPE14}
Moreover, the condition (\ref{Tc0-eq3}) on Eq.\ (\ref{Sigma_pp0^})
is known as {\it the Thouless criterion},\cite{Thouless60,NSR85}
which has been used to estimate superfluid transition temperatures of 
superconductors and atomic gases.\cite{NSR85,Haussmann93,Haussmann94,HRCZ07,CSTL05,KWO12,SPRSU18,BPE14}

In contrast, the second term of Eq.\ (\ref{U^s0}), which represents correlation effects of particle-particle bubbles on the
anomalous self-energy, has not been considered elsewhere at all due to the omission
of the mixing between the particle-particle and particle-hole channels through the emergence of 
the anomalous Green's function $F$.
Hence, there has not been any detailed microscopic study on
what happens after the condition (\ref{Tc0-eq3}) has been reached
with incorporating the second term on the right-hand side of Eq.\ (\ref{U^s0}).
Indeed, the divergence of the series of Fig.\ \ref{fig1}(a) has often been identified as 
signaling the superfluid transition whose transition temperature is determined by
the mean-field equation (\ref{Tc0-eq}), as assumed by Nozi\`eres and Schmitt-Rink\cite{NSR85}
and also seen in early studies on a different topic of ferromagnetic fluctuations\cite{IKK63,BS66,DE66}  where
the ferromagnetic phase is not considered at all.

Concerning ordered phases, there have been few studies by quantum field theory that 
focus on correlation effects of the particle-particle bubbles fully self-consistently beyond the mean-field treatment.
One of the exceptional ones is that by Haussmann {\it et al}.\ \cite{HRCZ07} on the attractive Hubbard model,
but the second term of Eq.\ (\ref{U^s0}) causing a $T_{\rm c}$ shift is missing
because they omit the particle-hole scattering process completely; see Appendix\ref{Appendix-Haussmann} on this point.
Moreover, existence of the divergence mentioned in the last paragraph of Sect.\ \ref{subsec:PE}
has been obscured by their procedure of replacing $U$ by the effective $s$-wave scattering length $a$.

Thus, the divergence mentioned in the last paragraph of Sect.\ \ref{subsec:PE} has been overlooked.
It should be noted in this context that correlation effects do have been investigated extensively by quantum field theory
in terms of cuprate superconductors, for example, mostly above $T_{\rm c}$ but some down below $T_{\rm c}$
based on the repulsive Hubbard model and related models.\cite{Yanase03,Kontani13,BSW89,BS89,PB94,MS94,MP94,LSGB95,DT95,DHS96,TM97,TM98}
However, those models do not yield any finite $T_{{\rm c}0}$ in the mean-field approximation
so that the divergence have not been encountered manifestly. Another point to be mentioned is that
the main focus there was on the correlation effects of the particle-hole bubbles called spin fluctuations
instead of the particle-particle bubbles of Fig.\ \ref{fig1}.

\subsection{Self-consistent treatment}

We now study the normal and superconducting phases on the same footing
based on the LW formalism for ordered phases.\cite{Kita11}
The grand thermodynamic potential $\Omega$ can be written generally as a functional of 
 the normal and anomalous Green's functions $(G,F)$ as $\Omega=\Omega[G,F]$.
Its equilibrium obeys the stationarity conditions
$\delta \Omega/\delta G=\delta \Omega/\delta F=0$, 
which form the Dyson-Gor'kov equation
\begin{align}
\hat{G}^{-1}(\vec{k}) \hat{G}(\vec{k}) =\hat{1},
\label{DG}
\end{align}
where $\hat{G}^{-1}$ and $\hat{G}$ are the Nambu matrices for the spin-singlet pairing given by
\begin{subequations}
\label{hatG^-1-hatG}
\begin{align}
\hat{G}^{-1}(\vec{k})=\begin{bmatrix} i\varepsilon_n-\xi_{\bf k}-\Sigma(\vec{k}) & -\Delta(\vec{k}) \\ -\Delta(\vec{k}) & 
i\varepsilon_n+\xi_{\bf k}+\Sigma(-\vec{k})\end{bmatrix},
\label{hatG^-1}
\end{align}
\begin{align}
\hat{G}(\vec{k})=\begin{bmatrix} G(\vec{k}) & F(\vec{k}) \\ F(\vec{k}) & -G(-\vec{k}) \end{bmatrix},
\label{hatG}
\end{align}
\end{subequations}
with $\Delta(-\vec{k})=\Delta(\vec{k})$ and $F(-\vec{k})=F(\vec{k})$.\cite{Kita11}
In addition, the self-energies $(\Sigma,\Delta)$ in Eq.\ (\ref{hatG^-1}) 
are defined in terms of a functional $\Phi=\Phi[G,F]$ that constitutes the LW functional by 
\begin{subequations}
\label{Sigma-Phi}
\begin{align}
\Sigma(\vec{k})=&\,\frac{1}{2T}\frac{\delta\Phi}{\delta G(\vec{k})},
\label{Sigma-Phi1}
\\
\Delta(\vec{k})=&\,\frac{1}{2T}\frac{\delta\Phi}{\delta F(\vec{k})}.
\label{Sigma-Phi2}
\end{align}
\end{subequations}
The key functional $\Phi$ consists of skeleton diagrams given as a power series in $U$;\cite{LW60,Baym62,DDM64,Kita96,HRCZ07,Kita11}
either cutting the series at a finite order or taking a partial sum provides us with a practical
approximation scheme to be solved self-consistently.

\begin{figure}[t]
\includegraphics[width=0.9\linewidth]{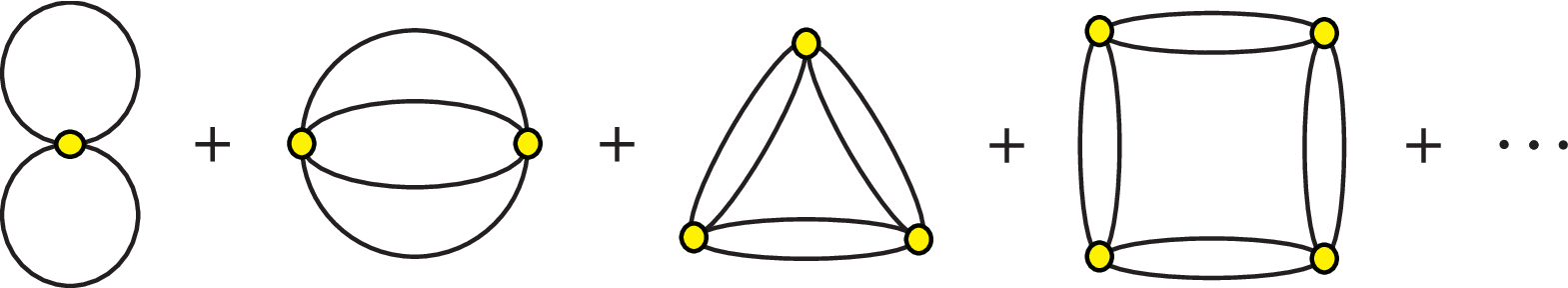}
\caption{\label{fig2}(Color online) Diagrammatic expression of $\Phi$ in the FLEX-S approximation \cite{Kita11}.
A small circle denotes the interaction vertex with particle-hole and spin degrees of freedom, and a line represents an element of $\hat{G}$.
}
\end{figure}

We here adopt the fluctuation-exchange approximation for superconductivity (FLEX-S approximation),\cite{Kita11}
whose $\Phi$ is given diagrammatically as Fig.\ \ref{fig2} and exhausts all the  anomalous processes
derivable from  the normal-state FLEX approximation\cite{Yanase03,Kontani13,BS89} 
consisting of the particle-particle and particle-hole scattering processes.
Specifically, incorporating only the first diagram in Fig.\ \ref{fig2} yields the mean-field BCS results with the Fermi-liquid corrections,\cite{Leggett75,SR83,Kita15} and higher-order diagrams represent correlation effects.
The differentiation of Eq.\ (\ref{Sigma-Phi}) corresponds to cutting and opening a line in Fig.\ \ref{fig2}.
The resulting self-energies $(\Sigma,\Delta)$ can be expressed in terms of the generalized bubble function
\begin{subequations}
\label{chi_AB}
\begin{align}
\chi_{AB}^{}(\vec{q})\equiv&\, -\frac{T}{{\cal N}_{\rm a}}\sum_{\vec{k}}A(\vec{k}+\vec{q})B(\vec{k}),
\label{chi_AB-def}
\end{align}
which satisfies
\begin{align}
\chi_{AB}^{}(\vec{q})\!=\!\chi_{BA}^{}(-\vec{q})\!=\! \chi_{\bar{A}\bar{B}}^{}(-\vec{q})
\label{chi_AB-symm}
\end{align}
\end{subequations}
with $\bar{B}(\vec{k})\equiv B(-\vec{k})$,
so as to incorporate the processes of Fig.\ \ref{fig1} naturally together with those with $F$'s exhaustively;
Eq.\ (\ref{chi_pp}) can be written in terms of Eq.\ (\ref{chi_AB-def}) as 
$$\chi_{\rm pp}^0(\vec{q})=-\chi_{G_0\bar{G}_0}(\vec{q}).$$
Specifically, let us introduce the functions
\begin{subequations}
\begin{align}
\chi_\pm^{}(\vec{q})\equiv &\,\chi_{GG}^{}(\vec{q})\pm \chi_{FF}^{}(\vec{q}) ,
\end{align}
\begin{align}
\underline{\chi}^{({\rm c})}(\vec{q})\equiv 
\begin{bmatrix}
\vspace{1mm}
\chi_{-}^{}(\vec{q})  & \sqrt{2}\chi_{GF}^{}(\vec{q}) & -\sqrt{2}\chi_{\bar{G}F}^{}(\vec{q})\\
\vspace{1mm}
\sqrt{2}\chi_{GF}^{}(\vec{q}) & -\chi_{G\bar{G}}^{}(\vec{q}) & -\chi_{FF}^{}(\vec{q}) \\
-\sqrt{2}\chi_{\bar{G}F}^{}(\vec{q}) & -\chi_{FF}^{}(\vec{q}) & -\chi_{\bar{G}G}^{}(\vec{q}) 
\end{bmatrix},
\label{chi^(0c)}
\end{align}
\end{subequations}
and the effective potentials
\begin{subequations}
\begin{align}
U_+^{\rm eff}(\vec{q})\equiv &\, U\biggl[\frac{U\chi_{+}(\vec{q})}{1-U\chi_{+}(\vec{q})}-\frac{1}{3}U\chi_{+}(\vec{q})\biggr] ,
\\
\underline{U}^{\rm eff}(\vec{q})\equiv &\,U\bigl[U\underline{\chi}^{({\rm c})}(\vec{q})\bigr]^2\bigl[\underline{1}+U\underline{\chi}^{({\rm c})}(\vec{q})\bigr]^{-1} .
\label{uU^eff}
\end{align}
\end{subequations}
Then we can express the self-energies as\cite{Kita11}
\begin{subequations}
\label{Sigma-Delta}
\begin{align}
\Sigma(\vec{k})  
 = &\, \frac{U {\cal N}_{\rm e}}{2{\cal N}_{\rm a}}
 +\frac{T}{{\cal N}_{\rm a}}\sum_{\vec{q}}U_{22}^{\rm eff}(\vec{q})G(-\vec{k}+\vec{q}) 
\notag \\
&\, +\frac{T}{2{\cal N}_{\rm a}}\sum_{\vec{q}}\bigl[3U_+^{\rm eff}(\vec{q})-U_{11}^{\rm eff}(\vec{q})\bigr] G(\vec{k}-\vec{q})
\notag \\
&\, -\frac{\sqrt{2}T}{{\cal N}_{\rm a}}\sum_{\vec{q}}U_{12}^{\rm eff}(\vec{q})F(\vec{k}-\vec{q}),
\label{Sigma}
\\
\Delta(\vec{k})  
 = &\, \frac{U T}{{\cal N}_{\rm a}}\sum_{\vec{k}'}F(\vec{k}')+\frac{T}{{\cal N}_{\rm a}}\sum_{\vec{q}}U_{23}^{\rm eff}(\vec{q})F(\vec{k}-\vec{q}) 
\notag \\
&\, +\frac{T}{2{\cal N}_{\rm a}}\sum_{\vec{q}}\bigl[3U_+^{\rm eff}(\vec{q})+U_{11}^{\rm eff}(\vec{q})\bigr] F(\vec{k}-\vec{q})
\notag \\
&\, -\frac{T}{{\cal N}_{\rm a}}\sum_{\vec{q}}U_{12}^{\rm eff}(\vec{q}) \frac{G(\vec{k}+\vec{q})+G(-\vec{k}+\vec{q})}{\sqrt{2}} ,
\label{Delta}
\end{align}
\end{subequations}
where $U_{ij}^{\rm eff}(\vec{q})$ denotes the $ij$ element of the matrix (\ref{uU^eff}),
and ${\cal N}_{\rm e}$ is the electron number defined by
\begin{align}
{\cal N}_{\rm e}= 2 T \sum_{\vec{k}}G(\vec{k})\, e^{i\varepsilon_n 0_+},
\label{N_e}
\end{align}
with $0_+$ denoting an infinitesimal positive constant.\cite{LW60}

\subsection{Normal self-energy and $T_{\rm c}$ equation\label{subsec:Tc-eq}}

Setting $F=0$ in Eq.\ (\ref{Sigma}) reproduces the 
self-energy of the normal-state FLEX approximation\cite{Yanase03,Kontani13,BS89} 
\begin{align}
\Sigma(\vec{k})=&\,\frac{U {\cal N}_{\rm e}}{2{\cal N}_{\rm a}}+\frac{UT}{{\cal N}_{\rm a}}\sum_{\vec{q}}\frac{U\chi_{G\bar{G}}^{}(\vec{q})}{1-U\chi_{G\bar{G}}^{}(\vec{q})}G(-\vec{k}+\vec{q})
\notag \\
&\,+\frac{3UT}{2 {\cal N}_{\rm a}}\sum_{\vec{q}}\frac{\bigl[U\chi_{GG}^{}(\vec{q})\bigr]^2}{1-U\chi_{GG}^{}(\vec{q})}G(\vec{k}-\vec{q})
\notag \\
&\,-\frac{UT}{2 {\cal N}_{\rm a}}\sum_{\vec{q}}\frac{\bigl[U\chi_{GG}^{}(\vec{q})\bigr]^2}{1+U\chi_{GG}^{}(\vec{q})}G(\vec{k}-\vec{q}),
\label{Sigma_n}
\end{align}
where the second term on the right-hand side corresponds to $\Sigma_{\rm pp}^0$ of 
Eq.\ (\ref{Sigma_pp0^}) with $G_0$ replaced by $G$.
Also linearizing Eq.\ (\ref{Delta}) with respect to $F(\vec{k})$ and replacing  $F(\vec{k})$ by Eq.\ (\ref{dF}),
we obtain the eigenvalue problem to determine the transition temperature that the equation
\begin{align}
-\frac{T}{{\cal N}_{\rm a}}\sum_{\vec{k}'}U_{\vec{k}\vec{k}'}^{\rm s}G(\vec{k}')G(-\vec{k}')\Delta(\vec{k}')=\epsilon \Delta(\vec{k})
\label{Tc-eq4}
\end{align}
with the pairing interaction
\begin{align}
U_{\vec{k}\vec{k}'}^{\rm s}
=&\,U\!\left[1-2x+\frac{3}{2}\frac{x}{1-x}-\frac{1}{2}\frac{x}{1+x}\right]_{x=U\chi_{GG}^{}(\vec{k}-\vec{k}')}
\notag \\
&\,-\frac{U^{2}T}{ {\cal N}_{\rm a}}\sum_{\vec{q}}\frac{[G(\vec{k}+\vec{q})+G(-\vec{k}+\vec{q})]G(\vec{k}'+\vec{q})}{\bigl[1+U\chi_{GG}^{}(\vec{q})\bigr]\bigl[1-U\chi_{G\bar{G}}^{}(\vec{q})\bigr]}
\label{Tc-FLEX-S-singlet}
\end{align}
has the eigenvalue $\epsilon=1$.
Note that Eq.\ (\ref{Tc0-eq4}) is reproduced from Eq.\ (\ref{Tc-FLEX-S-singlet}) by
setting the particle-hole bubble $\chi_{\rm ph}^{}\equiv\chi_{GG}^{}$ equal to $0$ except for the second-order contribution
and replacing $G$ by $G_0$.
Indeed, $\chi_{\rm ph}^{}$ becomes negligible in the weak-coupling limit of the attractive Hubbard model
compared with the particle-particle bubble $\chi_{\rm pp}^{}\equiv -\chi_{G\bar{G}}^{}$.

\subsection{Hidden BE singularity\label{subsec:HS}}

The hidden BE singularity of the attractive Hubbard model is defined 
by the condition that the equality
\begin{align}
U\chi_{G\bar{G}}(\vec{0})=1
\label{HS}
\end{align}
holds in the normal state.
It corresponds to the superconducting transition temperature $T_{{\rm c}0}^{\rm c}$ of 
{\it the correlated mean-field theory of superconductivity} where $G_0$ of the mean-field
Eq.\ (\ref{Tc0-eq2}) for the transition temperature $T_{{\rm c}0}$ is replaced by $G$ with correlations.
It may also be regarded as a generalization of the Thouless condition\cite{Thouless60,NSR85} to (i) make 
Green's function self-consistent and (ii) incorporate processes other than the particle-particle scattering process
in the calculation of the normal self-energy.
Concerning Eq.\ (\ref{HS}), 
it should be noted that $\chi_{G\bar{G}}({\bf q},0)$ in the zero Matsubara frequency branch is real,
as can be shown based on Eq.\ (\ref{chi_AB-def}) 
by using $\bigl[G({\bf k},i\varepsilon_n)\bigr]^*\!=\!G({\bf k},-i\varepsilon_n)$
and making a change of the summation variable $n\rightarrow-n$.

\begin{figure}[b]
\centering
\includegraphics[width=0.8\linewidth]{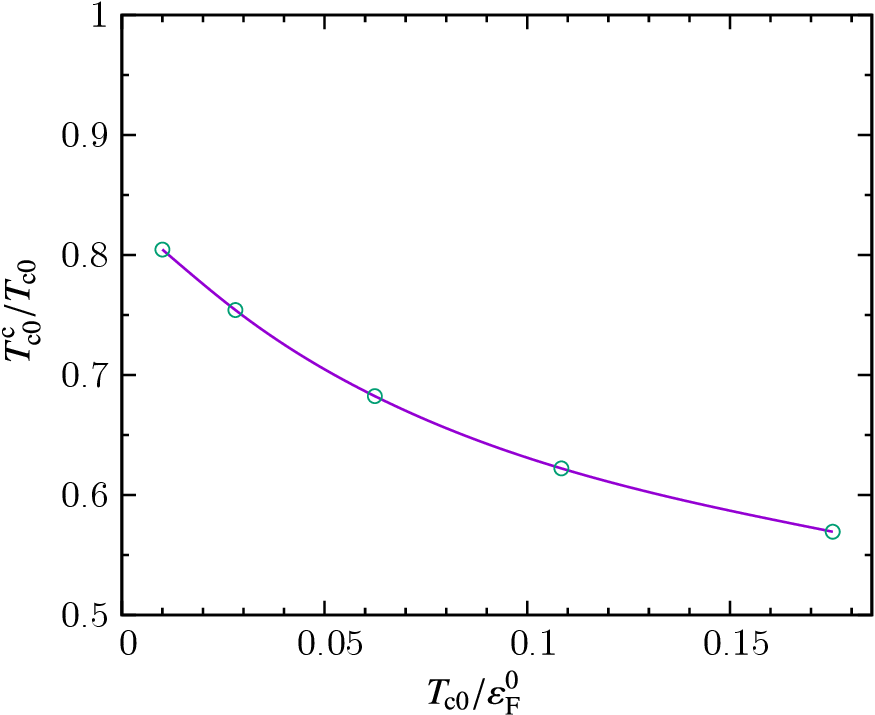}
\caption{\label{fig3}(Color online) Ratio $T_{{\rm c}0}^{\rm c}/T_{{\rm c}0}$ as a function of $T_{{\rm c}0}/\varepsilon_{\rm F}^0$
for the model band structure of $\xi_{\bf k}+\mu\propto k^2$ with $0\leq \xi_{\bf k}+\mu\leq 100\varepsilon_{\rm F}^0$,\cite{Kita24}
where $\varepsilon_{\rm F}^0$ is the non-interacting Fermi energy.
The line denotes the cubic-spline interpolation of the five circles obtained numerically
for $UN(0)/\varepsilon_{\rm F}^0=-0.07135,-0.077,-0.082,-0.086,-0.09$,
corresponding to $T_{{\rm c}0}=0.01,0.02791,0.06140,0.1074,0.1767$, respectively,
as calculated by incorporating the $T$ dependence of $\mu$ beyond Eq.\ (\ref{Tc0}).}
\end{figure}

Figure \ref{fig3} plots the ratio $T_{{\rm c}0}^{\rm c}/T_{{\rm c}0}$ as a function of $T_{{\rm c}0}$ 
calculated by using Eq.\ (\ref{Sigma_n}) for the normal self-energy and adopting the same band structure as the previous study,\cite{Kita24} 
for which $T_{{\rm c}0}$ in the weak coupling can be expressed analytically following the standard procedure\cite{BCS57,Bogoliubov59,deGennes66,AGD63,Parks69,Kita15} as
\begin{align}
T_{{\rm c}0}=C\varepsilon_{\rm F}^0\,\exp\left[\frac{\varepsilon_{\rm F}^0}{UN(0)}\right]
\label{Tc0}
\end{align}
where $\varepsilon_{\rm F}^0$ is the non-interacting Fermi energy,
$N(0)$ is the density of states per spin at $\varepsilon_{\rm F}^0$, and $C$ is a constant given by
\begin{align}
C \equiv \frac{8 e^{\gamma-2}}{\pi}\exp\left(\sqrt{\frac{\varepsilon_{\rm c}}{\varepsilon_{\rm F}^0}}-\frac{1}{2}\ln\left|\frac{1+\sqrt{\varepsilon_{\rm F}^0/\varepsilon_{\rm c}}}{1-\sqrt{\varepsilon_{\rm F}^0/\varepsilon_{\rm c}}}\right|\,\right)
\end{align}
in terms of the Euler constant $\gamma =0.577$ and the upper limit of the band energy $\varepsilon_{\rm c}=100\varepsilon_{\rm F}^0$.\cite{Kita24}
One can see from Fig.\ \ref{fig3} that (i) $T_{{\rm c}0}^{\rm c}/T_{{\rm c}0}<1$ holds generally and (ii) the ratio apparently does not reach $1$ for $T_{{\rm c}0}\rightarrow 0$.
Indeed, it is shown in Appendix\ref{Appendix-C} that $T_{{\rm c}0}^{\rm c}/T_{{\rm c}0}$ approaches 
$1$ quite slowly with the logarithmic dependence
\begin{align}
\frac{T_{{\rm c}0}^{\rm c}}{T_{{\rm c}0}}= 1-\frac{a}{[\ln (C\varepsilon_{\rm F}^0/T_{{\rm c}0})]^2}+\cdots,
\label{T_c0^c-WC}
\end{align}
where $a$ is a constant.
Thus,  the mean-field transition temperature is lowered considerably even in the weak-coupling region
by incorporating correlations in the normal state, 
implying that the correlations change the nature of the superconducting transition quantitatively and substantially. 
This will be a feature qualitatively common to any band structure of the attractive Hubbard model;
we have confirmed the statement for the quadratic dispersion by changing $\varepsilon_{\rm c}$.

Now, Eq.\ (\ref{HS}) results from Eq.\ (\ref{Tc-eq4}) with $\epsilon=1$ by approximating $U_{\vec{k}\vec{k}'}^{\rm s}\approx U$.
However, the presence of the higher-order terms in Eq.\ (\ref{Tc-FLEX-S-singlet}) generally causes a shift of the 
superconducting transition temperature $T_{\rm c}$ from the value $T_{{\rm c}0}^{\rm c}$ evaluated by Eq.\ (\ref{HS}),
specifically to a lower value for the attractive Hubbard model.
Indeed, our numerical study performed at $T>T_{{\rm c}0}^{\rm c}$ shows that 
the eigenvalue $\epsilon$ ($<1$) is lowered due to the correlation contribution 
of $U_{\vec{k}\vec{k}'}^{\rm s}-U\neq 0$ in Eq.\ (\ref{Tc-FLEX-S-singlet}).
This shift of the superconducting transition temperature $T_{\rm c}$ from the correlated mean-field value $T_{{\rm c}0}^{\rm c}$
is an intrinsic feature caused by correlation effects in the anomalous self-energy $\Delta(\vec{k})$.

However, passing $T_{{\rm c}0}^{\rm c}$ down to lower temperatures makes any self-consistent calculation impossible 
due to the divergence of the self-energy (\ref{Sigma_n}) 
through entering the region of $U\chi_{G\bar{G}}(\vec{0})>1$;
see the argument in the last paragraph of Sect.\ \ref{subsec:PE} on this point.
The condition $U\chi_{G\bar{G}}(\vec{0})>1$ is also unacceptable physically 
since it implies that the argument of the logarithm in
\begin{align}
\Phi_{\rm pp}\equiv T\sum_{\vec{q}} \Bigl\{\ln\left[1-U\chi_{\rm G\bar{G}}(\vec{q})\right]+U\chi_{\rm G\bar{G}}(\vec{q})\Bigr\},
\label{Omega_pp}
\end{align}
which represents the contribution of the particle-particle bubbles to $\Phi$,
becomes negative in the $\omega_\ell=0$ branch over a finite region of ${\bf q}$.
In this context, it should be noted that presence of the logarithmic contribution composed of the one-loop bubbles
has been well established physically in a different context of the screening of the electron gas
in terms of the particle-hole bubbles.\cite{Macke50,BP53,GB57}

\section{Removal of Divergence and Its Consequences}

\subsection{Physical condition to remove divergence}

The arguments of Sect.\ \ref{subsec:Tc-eq} and \ref{subsec:HS} indicate that
the superconducting transition temperature $T_{\rm c}$ will be lowered from the correlated mean-field value
$T_{{\rm c}0}^{\rm c}$ determined by the one-loop condition $U\chi_{G\bar{G}}(\vec{0})=1$
due to correlation effects in the anomalous self-energy.
On the other hand, Eq.\ (\ref{Omega_pp}) tells us that $T_{{\rm c}0}^{\rm c}$ is a definite singular point
of the grand thermodynamic potential. 
Hence, $T_{{\rm c}0}^{\rm c}$ should remain as a singular point distinct from $T_{\rm c}$, which we call the hidden BE singularity,
below which the standard LW formalism cannot yield any physical solution.

This difficulty is caused by entering the region of $U\chi_{G\bar{G}}(\vec{0})>1$,
which is unphysical and impossible according to Eq.\ (\ref{Omega_pp}).
What must be happening {\it physically} is that the system adjusts itself 
so that $U\chi_{G\bar{G}}(\vec{0})$ keeps the value $1$ once it has been reached,
similarly as the chemical potential of ideal Bose gases upon passing down through the Bose--Einstein condensation temperature.\cite{Kita15}
The condition can be incorporated {\it mathematically} in the LW formalism by adding 
to the $\Phi$ functional the following term that vanishes for $U\chi_{G\bar{G}}^{}(\vec{0})=1$:
\begin{align}
\varDelta\Phi_\lambda \equiv \lambda {\cal N}_{\rm e}\bigl[1-U\chi_{G\bar{G}}^{}(\vec{0})\bigr],
\label{Phi_lambda^n}
\end{align}
where $\lambda$ is the Lagrange multiplier.
Performing the differentiation of Eq.\ (\ref{Sigma-Phi1}) with Eq.\ (\ref{Phi_lambda^n}) added to $\Phi$, 
we obtain the expression of the self-energy for the pinned phase of $U\chi_{G\bar{G}}^{}(\vec{0})=1$ as
\begin{align}
\Sigma_{\lambda}(\vec{k}) =\Sigma(\vec{k})+\lambda\bar{n}U G(-\vec{k}) 
\label{dSigma_lambda}
\end{align}
where $\Sigma(\vec{k})$ is given by Eq.\ (\ref{Sigma_n}) and $\bar{n}\!\equiv\! {\cal N}_{\rm e}/{\cal N}_{\rm a}$
denotes the number of electrons per site.
The second term stemming from Eq.\ (\ref{Phi_lambda^n})
is characteristic of the pinned phase.
It has an unusual structure called {\it one-particle reducible} (1PR), i.e., the structure that becomes 
disconnected upon cutting a line,\cite{CJT74}
which originates from the two-particle-reducible structure of Eq.\ (\ref{Phi_lambda^n}).
Thus, $T_{{\rm c}0}^{\rm c}$ is identified in our theory with the critical temperature $T_{\rm 1PR}$ at which the 
one-particle-reducible structure emerges in the self-energy.
Using Eq.\ (\ref{dSigma_lambda}), we can calculate properties of the normal phase from $T_{\rm 1PR}=T_{{\rm c}0}^{\rm c}$
down to $T_{\rm c}$.

To describe the superconducting phase, 
we need to extend Eq.\ (\ref{Phi_lambda^n}) so as to incorporate the contribution of the $F$ function.
The condition $U\chi_{G\bar{G}}^{}(\vec{0})=1$ for the normal phase can be generalized 
to the superconducting phase in terms of Eq.\ (\ref{chi^(0c)}) as\cite{Kita24} 
\begin{align}
\det\left[\underline{1}+U\underline{\chi}^{({\rm c})}(\vec{0})\right]=0,
\label{det(1+U*Chi^c)=0}
\end{align}
where $\underline{1}$ is the $3\times 3$ unit matrix.
A straightforward calculation of the determinant shows that this equality can be simplified to
\begin{align}
U\bigl[\chi_{G\bar{G}}(\vec{0})+\chi_{FF}(\vec{0})\bigr]=1,
\label{pin-super}
\end{align}
which is a direct extension of $U\chi_{G\bar{G}}^{}(\vec{0})=1$ for the normal phase.
Its validity is also supported by our previous study that ${\lambda}_G={\lambda}_F=\tilde{\lambda}_G=0$ holds
numerically for the parameters given below Eq.\ (17) of Ref.\ \onlinecite{Kita24}
when Eq.\ (\ref{det(1+U*Chi^c)=0}) is satisfied.
Hence, Eq.\ (\ref{Phi_lambda^n}) for the normal phase is generalized to
\begin{align}
\varDelta\Phi_\lambda=\lambda {\cal N}_{\rm e}\left[1-U\chi_{G\bar{G}}(\vec{0})-U\chi_{FF}(\vec{0})\right] .
\label{Phi_lambda}
\end{align}
Performing the differentiations of Eq.\ (\ref{Sigma-Phi}) with Eq.\ (\ref{Phi_lambda}) added to $\Phi$, 
we obtain the self-energies for the pinned phase of satisfying Eq.\ (\ref{pin-super}) as
\begin{subequations}
\label{hatSigma_lambda}
\begin{align}
\Sigma_{\lambda}(\vec{k}) =&\,\Sigma(\vec{k})+\lambda\bar{n}U G(-\vec{k}) ,
\label{Sigma_lambda}
\\
\Delta_{\lambda}(\vec{k}) =&\,\Delta(\vec{k})+\lambda\bar{n}U F(\vec{k}) .
\label{Delta_lambda}
\end{align}
\end{subequations}
where $\Sigma$ and $\Delta$ on the right-hand sides are given by Eqs.\ (\ref{Sigma}) and (\ref{Delta}), respectively.

\subsection{Expression of Green's function}

From now on we express the $2\times 2$ matrix Green's function for $\lambda\neq 0$ and its inverse
as $\hat{G}_\lambda(\vec{k})$ and $\hat{G}_\lambda^{-1}(\vec{k})$, respectively.
It follows from Eqs.\ (\ref{hatG^-1-hatG}) and (\ref{hatSigma_lambda}) that $\hat{G}_\lambda^{-1}$ and $\hat{G}^{-1}$ are connected by
\begin{align}
\hat{G}_\lambda^{-1}(\vec{k})=\hat{G}^{-1}(\vec{k})-\bar\lambda \hat{G}_\lambda(-\vec{k}),
\label{hG_lambda^-1}
\end{align}
with
\begin{align}
\bar\lambda\equiv \lambda\bar{n}U .
\end{align}
Our purpose here is to show that $\hat{G}_\lambda(\vec{k})$ obeying Eq.\ (\ref{hG_lambda^-1}) can be expressed
in terms of $\hat{G}(\vec{k})\equiv\bigl[\hat{G}^{-1}(\vec{k})]^{-1}$ as
\begin{align}
\hat{G}_\lambda(\vec{k})=\phi_\lambda(\vec{k})\hat{G}(\vec{k}),
\label{G_lambda-sol}
\end{align}
where $\phi_\lambda(\vec{k})$ is a scalar function defined by
\begin{align}
\phi_\lambda(\vec{k})\equiv \frac{2}{1+\sqrt{1+4\bar\lambda/{\cal D}(\vec{k})}},
\end{align}
with
\begin{align}
{\cal D}(\vec{k})\equiv &\,\det \hat{G}^{-1}(\vec{k}) .
\label{D(k)-def}
\end{align}
Green's function of the normal phase with $\lambda\neq 0$ is obtained from Eq.\ (\ref{G_lambda-sol}) by setting $F=0$.

The proof of Eq.\ (\ref{G_lambda-sol}) proceeds as follows.
Multiplying $\hat{G}_\lambda^{-1}(-\vec{k})= \hat{G}^{-1}(-\vec{k})-\bar\lambda \hat{G}_\lambda(\vec{k})$
from the left-hand side of Eq.\ (\ref{hG_lambda^-1}), we obtain
\begin{align}
\hat{G}^{-1}(-\vec{k})\hat{G}^{-1}_\lambda(\vec{k})=\hat{G}^{-1}(-\vec{k})\hat{G}^{-1}(\vec{k})-\bar\lambda \hat{G}_\lambda(\vec{k})\hat{G}^{-1}(\vec{k}) .
\label{G_lambda^-1(k)-eq}
\end{align}
It follows from Eq.\ (\ref{hatG^-1}) that $\hat{G}^{-1}(-\vec{k})\hat{G}^{-1}(\vec{k})$ is proportional to the unit matrix $\hat{1}$ as
\begin{align}
\hat{G}^{-1}(-\vec{k})\hat{G}^{-1}(\vec{k})=-{\cal D}(\vec{k})\hat{1},
\label{G^-1(k)G^-1(-k)}
\end{align}
where ${\cal D}(\vec{k})$ is defined by Eq.\ (\ref{D(k)-def}). 
Let us operate $\hat{G}_\lambda^{-1}(\vec{k})$ from the left-hand side of Eq.\ (\ref{G_lambda^-1(k)-eq}),
substitute Eq.\ (\ref{G^-1(k)G^-1(-k)}), and multiply the resulting equation by $G^{-1/2}(-\vec{k})$ from both sides.
We thereby obtain 
\begin{align}
[\hat{\cal G}^{-1}_\lambda(\vec{k})]^2=-{\cal D}(\vec{k})
\hat{\cal G}^{-1}_\lambda(\vec{k})+\bar\lambda {\cal D}(\vec{k}) \hat{1},
\label{G^-1-eq2}
\end{align}
with
\begin{align}
\hat{\cal G}_\lambda^{-1}(\vec{k})\equiv \hat{G}^{-1/2}(-\vec{k}) \hat{G}^{-1}_\lambda(\vec{k})\hat{G}^{-1/2}(-\vec{k}).
\label{hatCalG}
\end{align}
Equation (\ref{G^-1-eq2}) can be transformed into
\begin{align*}
\left[\hat{\cal G}^{-1}_\lambda(\vec{k})+\frac{1}{2}{\cal D}(\vec{k})\hat{1}\right]^2=\frac{[{\cal D}(\vec{k})]^2}{4}\left[1+4\bar\lambda {\cal D}^{-1}(\vec{k}) \right]\hat{1},
\end{align*}
which can be solved easily as
\begin{align}
\hat{\cal G}^{-1}_\lambda(\vec{k})=\frac{-{\cal D}(\vec{k})}{2}\left[1+\sqrt{1+4\bar\lambda /{\cal D}(\vec{k})}\right]\hat{1},
\label{calG-eq}
\end{align}
where we have adopted the $+$ sign in front of the square root to reproduce the $\bar\lambda=0$ result.
Let us substitute Eq.\ (\ref{hatCalG}) in Eq.\ (\ref{calG-eq}) and 
solve the resulting equation in terms of $\hat{G}_\lambda(\vec{k})$
by using the relation $\hat{G}(-\vec{k}){\cal D}(\vec{k})=-\hat{G}^{-1}(\vec{k})$ from Eq.\ (\ref{G^-1(k)G^-1(-k)}). 
We thereby obtain Eq.\ (\ref{G_lambda-sol}).

\subsection{Changes in the density of states}

The factor $\phi_\lambda$ in Eq.\ (\ref{G_lambda-sol}) causes a substantial change in the single-particle density of states defined by
\begin{align}
D(\varepsilon)\equiv \frac{2}{{\cal N}_{\rm a}}\sum_{{\bf k}}\frac{-1}{\pi}{\rm Im}\,G_\lambda({\bf k},i\varepsilon_n\rightarrow \varepsilon+i0_+),
\label{D(e)-def}
\end{align}
where Im denotes taking the imaginary part.
To see this qualitatively, we consider the non-interacting normal-state Green's function 
$G_0^{-1}(\vec{k})\equiv i\varepsilon_n-\xi_{\bf k}$. 
The corresponding $G_{0\lambda}(\vec{k})$ can be obtained easily as
\begin{align}
G_{0\lambda}(\vec{k})=\frac{2}{i\varepsilon_n-\xi_{\bf k}}\left[1+\sqrt{\frac{(i\varepsilon_n)^2-{\cal E}_{\bf k}^2}{(i\varepsilon_n)^2-\xi_{\bf k}^2}}\,\right]^{-1}
\end{align}
with ${\cal E}_{\bf k}\equiv \sqrt{\xi_{\bf k}^2-4\bar\lambda}$. 
We focus on the case $\bar\lambda\equiv \lambda\bar{n}U <0$ based on our previous numerical study.\cite{Kita24}
Through the analytic continuation $i\varepsilon_n\rightarrow \varepsilon+i0_+$,  we obtain the spectral function
$\rho_{0\lambda}({\bf k},\varepsilon)\equiv \frac{-1}{\pi}{\rm Im}\,G_{0\lambda}({\bf k}, \varepsilon+i0_+)$ as
\begin{align}
\rho_{0\lambda}({\bf k},\varepsilon)=\left\{\begin{array}{cl}\vspace{1mm}
\displaystyle \!\!\!\frac{\sqrt{({\cal E}_{\bf k}^2-\varepsilon^2)(\varepsilon^2-\xi_{\bf k}^2)}}{-2\pi \bar\lambda |\varepsilon-\xi_{\bf k}|} & 
:\,|\xi_{\bf k}|\leq |\varepsilon| \leq {\cal E}_{\bf k} \\
0 & :\,\mbox{otherwise}\end{array}\right. \!\! .
\label{rho_0}
\end{align}
Thus, the sharp peak $\delta(\varepsilon-\xi_{\bf k})$ for $\lambda=0$ is split and broadened into 
the two regions $-{\cal E}_{\bf k}\leq \varepsilon \leq -|\xi_{\bf k}|$ and $|\xi_{\bf k}|\leq \varepsilon \leq {\cal E}_{\bf k}$,
leaving a weaker divergence at the original position $\varepsilon=\xi_{\bf k}$.
This change of the spectral function gives rise to a decrease in the density of states around $\varepsilon=0$.
Indeed, using Eq.\ (\ref{rho_0}) in Eq.\ (\ref{D(e)-def}) and approximating ${\cal E}_{\bf k}^2-\varepsilon^2\approx -4\bar\lambda$
for $|\xi_{\bf k}|\leq |\varepsilon|\ll -4\bar\lambda$,
we find that the density of states evaluated by $G_{0\lambda}(\vec{k})$ for $|\varepsilon|\ll -4\bar\lambda$
vanishes towards zero energy with the characteristic V-shape structure
\begin{align}
D_0(\varepsilon)\approx \frac{2N(0)}{\sqrt{-\bar\lambda}}|\varepsilon| ,
\label{D_0}
\end{align}
where $N(0)$ denotes the non-interacting density of states per spin and per site at $\varepsilon_{\rm F}^0$.
Thus, the factor $\phi_\lambda$ brings a substantial reduction in the normal-state density of states
near the excitation threshold,
which we call {\it a pseudogap behavior}.

The superconducting phase can be analyzed similarly based on Green's function of the BCS theory\cite{AGD63,Parks69,Kita15}
\begin{align}
G_{\rm BCS}(\vec{k})\equiv \frac{u_{\bf k}^2}{i\varepsilon_n-E_{\bf k}}+\frac{v_{\bf k}^2}{i\varepsilon_n+E_{\bf k}},
\end{align}
with $E_{\bf k}\equiv \sqrt{\xi_{\bf k}^2+\Delta^2}$, $u_{\bf k}^2=\frac{1}{2}\bigl(1+\frac{\xi_{\bf k}}{E_{\bf k}}\bigr)$, 
and $v_{\bf k}^2=\frac{1}{2}\bigl(1-\frac{\xi_{\bf k}}{E_{\bf k}}\bigr)$.
Repeating the above argument on the normal phase,
we can conclude that
each of the sharp peaks $u_{\bf k}^2\, \delta(\varepsilon-E_{\bf k})$ and $v_{\bf k}^2 \,\delta(\varepsilon+E_{\bf k})$
at $\varepsilon=\pm E_{\bf k}$ is split and broadened into the two regions
 $-{\cal E}_{\bf k}\leq \varepsilon \leq -E_{\bf k}$ and $E_{\bf k} \leq \varepsilon \leq {\cal E}_{\bf k}$
 with ${\cal E}_{\bf k}\equiv \sqrt{E_{\bf k}^2-4\bar\lambda}$.
It hence follows that the factor $\phi_\lambda$ only causes some modification of $D(\varepsilon)$ for 
 $|\varepsilon|>\Delta$ but does not affect $D(\varepsilon)$ for 
 $|\varepsilon|<\Delta$ at all.

\section{Numerical Results}

\begin{figure}[t]
\centering
\hspace{7mm}\includegraphics[width=0.8\linewidth]{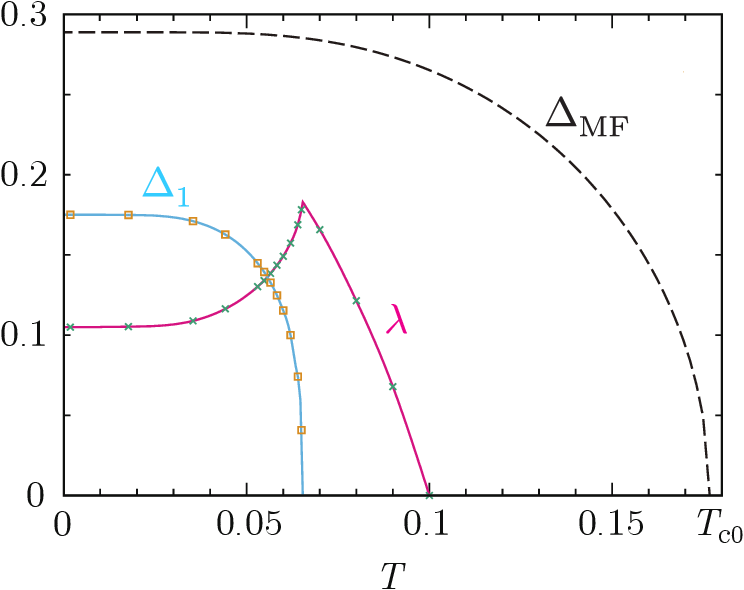}
\caption{\label{fig4}(Color online) Temperature dependences of the Lagrange multiplier $\lambda$
and the first-order anomalous self-energy $\Delta_1^{}$ in comparison with that of the mean-field energy gap $\Delta_{\rm MF}^{}$.
}
\end{figure}

We have performed a fully self-consistent numerical calculation of the 
density of states (\ref{D(e)-def})
from $T>T_{\rm 1PR}=T_{{\rm c}0}^{\rm c}$ down to $T=0$
 for a low-density  attractive Hubbard model considered previously.\cite{Kita24}
Specifically, the band structure is given by $\xi_{\bf k}=k^2-\mu$ with $0\leq k\leq 10$ in units of $2m\!=\!k_{\rm F}\!=\!\varepsilon_{\rm F}^0=1$, 
where $m$ is the electron mass, $k_{\rm F}$ is the Fermi momentum, and $\varepsilon_{\rm F}^0=\mu_0(T=0)$ denotes the non-interacting Fermi energy.
Temperature dependences of the basic parameters were obtained as Fig.\ \ref{fig4}
at an intermediate coupling of $UN(0)=-0.09$,\cite{Kita24} where $\Delta_{\rm MF}$ and 
$T_{{\rm c}0}\!=\!0.1767$ are the mean-field results on the energy gap and transition temperature,
$\lambda$ is the prefactor of the 1PR contribution to the self-energy in Eq.\ (\ref{hG_lambda^-1}),
and $\Delta_1^{}$ is the energy gap evaluated by the first term 
on the right-hand side of Eq.\ (\ref{Delta}) in terms of the fully self-consistent $F$.
The critical temperatures $(T_{\rm c},T_{\rm 1PR})$ satisfy $T_{\rm c}<T_{\rm 1PR}<T_{{\rm c}0}$
with $T_{\rm c}=0.0654$ and $T_{\rm 1PR}=0.100$.
Combined with Fig.\ \ref{fig3} on $T_{{\rm c}0}^{\rm c}=T_{\rm 1PR}$, we may assume that
the qualitative features of Fig.\ \ref{fig4} remain the same over a wide range from the intermediate- to the weak-coupling regions
irrespective of the band structure.

\begin{figure}[b]
\centering
\includegraphics[width=0.8\linewidth]{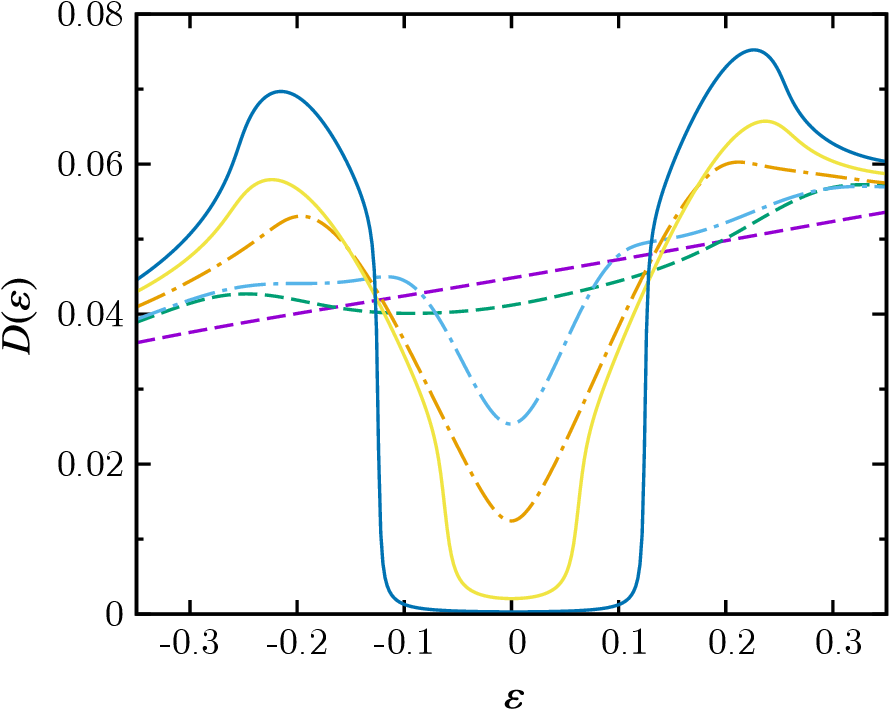}
\caption{\label{fig5}(Color online) The density of states $D(\varepsilon)$ calculated at $T=0.20$, $0.10$, $0.096$, $0.08$, $0.064$, $0.0353$ 
from top to bottom at $\varepsilon=0$. The first two dashed lines are in the normal state, the V-shaped dash-dotted curves 
are in the pseudogap phase of $\lambda>0$, and the last two curves are in the superconducting phase.}
\end{figure}

Figure \ref{fig5} plots the density of states calculated at six different temperatures
obtained by a numerical analytic continuation 
of the self-energies based on the six-point Pad\'e approximant\cite{VS77} for $\Sigma(\vec{k})$ and five-point one for $\Delta(\vec{k})$
on the right-hand sides of Eqs.\ (\ref{Sigma_lambda}) and (\ref{Delta_lambda}), respectively.
The two dashed curves are in the normal state of $T=0.20$, $0.10$; 
the dash-dotted ones are in the pseudogap phase of $\lambda>0$ at $T=0.096$, $0.08$, which exhibit the characteristic V-shape structure 
of Eq.\ (\ref{D_0}); the two full lines are in the superconducting phase
at $T= 0.064$, $0.0353$ showing the U-shape structure at low energies as expected for the $s$-wave energy gap.
Thus, the distinctions among the normal, pseudogap, and superconducting phases are predicted to be 
detectable for the present $s$-wave pairing by observing the density of states.
More detailed features of these curves are summarized as follows. 
First, a slight decrease near $\varepsilon=0$ at $T=0.10$ from the $T=0.20$ result is caused by 
the combined effects of (i)
the change of slope $\partial{\rm Re}\Sigma({\bf k},i0_+)/\partial k\bigr|_{k=k_{\rm F}}$ and
(ii) the increase of $-\frac{1}{\pi}{\rm Im}\Sigma({\bf k}_{\rm F},\varepsilon+i0_+)$ around $\varepsilon=0$
upon approaching $T_{\rm 1PR}$.
Second, the density of states in the pseudogap phase remains finite at $\varepsilon=0$
unlike Eq.\ (\ref{D_0}) of the non-interacting case due to $-\frac{1}{\pi}{\rm Im}\Sigma({\bf k},\varepsilon+i0_+)>0$.
Third,  temperature variation of the superconducting energy gap estimated by Fig.\ \ref{fig5}
almost follows that of $\Delta_1$ in Fig.\ \ref{fig4}.
In addition, the finite offset around $\varepsilon=0$, which becomes manifest at $T=0.064$,
is also caused by $-\frac{1}{\pi}{\rm Im}\Sigma({\bf k},i0_+)>0$.

Thus, the present theory predicts appearance of a pseudogap in the single-particle density of states
upon passing down the Bose-Einstein singularity due to the emerging 1PR structure of Eq.\ (\ref{dSigma_lambda}).
This 1PR structure may be regarded alternatively as an emerging pole in the self-energy,
the possibility of which has been discussed rather phenomenologically 
in terms of cuprate superconductors.\cite{Norman07}
Equation (\ref{dSigma_lambda}) may be regarded as providing a definite microscopic mechanism 
for the emerging pole in the self-energy with the threshold temperature $T_{\rm 1PR}$.
The pseudogap behaviors have been observed in strongly correlated superconductors\cite{TS99, Kordyuk15} and atomic gases\cite{SGJ08,GSDJPPS10}
above the superfluid transition temperatures.
The results on atomic gases have been analyzed in terms of the attractive Hubbard model
using various techniques such as the $T$-matrix approximation,\cite{HRCZ07,CSTL05,KWO12,BPE14} dynamical mean-field theory (DMFT),\cite{PB15,Sakai15} and quantum Monte Carlo (QMC).\cite{MIB11}
One of the DMFT results has been analyzed successfully in terms of an analytic model
that yields a pole in the self-energy.\cite{Sakai15}
The present study predicts that a pseudogap phase is present above the superfluid phase 
of the attractive Hubbard model from the very weak-coupling region over a wide range of the interaction strength
in a way distinguishable by the characteristic V-shape structure in the density of states
near zero energy.

\section{Summary and Discussions}

The attractive Hubbard model in three dimensions has been studied based on quantum field theory of ordered phases\cite{Kita24}
over a wide range of the interaction strength in the weak-coupling region.
The FLEX-S approximation adopted here has a definite advantage over the $T$-matrix approximation used extensively in
the literature\cite{Yanase03,NSR85,Haussmann93,Haussmann94,HRCZ07,CSTL05,KWO12,SPRSU18,BPE14}
 in that: (i) the contribution of particle-hole bubbles are included besides that of the particle-particle bubbles 
in the normal state; (ii) all the anomalous processes of superconductivity derivable from the two normal processes
are incorporated naturally.
This approximation collects all the leading-order diverging contributions in the weak-coupling
region, leaving no omitted terms that may affect the result qualitatively nor quantitatively.
It is thereby shown that correlation effects on the anomalous self-energy split
the superconducting transition temperature $T_{\rm c}$ down below
the value $T_{{\rm c}0}^{\rm c}$ determined by the generalized Thouless criterion, 
leaving $T_{{\rm c}0}^{\rm c}$ as a distinct BE singularity,
below which the self-energy encounters a divergence to make the standard self-consistency procedure of
the LW formalism useless.
On the basis of a physical consideration to stop the divergence,
this temperature $T_{{\rm c}0}^{\rm c}$ is identified here 
with the threshold temperature $T_{\rm 1PR}$ of marking emergence of a 1PR  structure in the self-energy,
which results in a pseudogap in the single-particle density of states with a characteristic V-shape structure around zero energy.
How the 1PR structure affects thermodynamic quantities such as specific heat remains 
to be clarified in the future.
The mechanism of the pseudogap formation presented here may also explain those observed in high-$T_{\rm c}$ superconductors\cite{TS99, Kordyuk15}
by identifying the relevant BE singularity, which may be an antiferromagnetic one rather than the superconducting one, for example.

Theoretically, it is important to establish 
the phase diagram of the attractive Hubbard model in the weak-coupling region. 
Completing the phase diagram for a finite band in terms of the bare interaction parameter $U$
will clarify whether the divergence mentioned below Eq.\ (\ref{subsec:PE}) 
is a real physical entity to yield the pseudogap phase as predicted here
or merely an artifact originating from the limitation of quantum field theory.
To this end, theoretical studies on the attractive Hubbard model
by other methods, such as DMFT or QMC, are desirable in the intermediate coupling region
where they can yield reliable results, since the phase diagram will remain invariant by passing
from the weak- to intermediate-coupling region.
It should be noted in this context that one of the previous DMFT studies\cite{Sakai15} already detects emergence of a pole
in the self-energy of the pseudogap phase, in agreement with the present result.

One of the effects not incorporated in the present study is the
superconducting fluctuations, which the present study show distinct from the
particle-particle scattering process of Fig.\ \ref{fig1}(a), contrary to the widely accepted viewpoint.\cite{Yanase03,NSR85,Haussmann93,Haussmann94,HRCZ07,CSTL05,KWO12,SPRSU18,BPE14}
To collect them systematically, we may be required to adopt some other technique such as the Stratonovich-Hubbard
transformation;\cite{Stratonovich57,Hubbard59} the issue here is whether $T_{0{\rm c}}^{\rm c}$ still remains as a definite singularity distinct from $T_{\rm c}$ or not,
but it is difficult to imagine that the fluctuations do have so large an effect in the weak-coupling region
to make $T_{0{\rm c}}^{\rm c}$ identical with $T_{\rm c}$.
Thus, quantum field theory of ordered phases still has a lot to be explored and developed.

\begin{acknowledgments}
This work was supported in part by JSPS KAKENHI Grant Number JP20K03848.
\end{acknowledgments}

\appendix

\section{Mean-Field Equation for $T_{{\rm c}0}$\label{AppA}}

We here show that, for ordered phases that are realized continuously, the mean-field equations to determine transition temperatures can be expressed generally in terms of a product of two Green's functions.
First, consider the case of superconductivity, which is characterized by the emergence of the anomalous Green's function $F$
and the corresponding self-energy $\Delta$ called {\it energy gap}.
In the mean-field approximation, they are calculated self-consistently by the so-called gap equation\cite{AGD63,Parks69,Kita15}
\begin{align}
\Delta =U\frac{T}{{\cal N}_{\rm a}}\sum_{\vec{k}} F(\vec{k}).
\label{Tc0-eqA}
\end{align}
On the other hand, $F$ obeys the Dyson-Gor'kov equation (\ref{DG}),
whose first-order variation can be written as
\begin{align*}
\hat{G}^{-1}(\vec{k}) \delta\hat{G}(\vec{k}) +[\delta\hat{G}^{-1}(\vec{k})] \hat{G}(\vec{k}) =\hat{0},
\end{align*}
i.e., 
\begin{align}
\delta\hat{G}(\vec{k}) \!=\!-\hat{G}(\vec{k}) \bigl[\delta\hat{G}^{-1}(\vec{k})\bigr] \hat{G}(\vec{k}).
\label{deltaG}
\end{align}
By identifying $F$ and $\Delta$ in Eq.\ (\ref{hatG^-1-hatG}) with the first-order variations of Eq.\ (\ref{deltaG})
from the normal-state $\hat{G}$ with $F=0$, we obtain $\delta F$ that is first order in $\Delta$ as
\begin{align}
\delta F(\vec{k})=-G(\vec{k})G(-\vec{k}) \Delta .
\label{dF}
\end{align}
Substituting Eq.\ (\ref{dF}) in Eq.\ (\ref{Tc0-eqA}) with $G$ replaced by $G_0$
and setting $T=T_{{\rm c}0}$, we obtain Eq.\ (\ref{Tc0-eq2}).

We can see from the above derivation that it is Eq.\ (\ref{deltaG}) for the linear emergence of a new element
that is responsible for the two $G$'s in Eq.\ (\ref{Tc0-eq2}),
implying that the feature should be common to all continuous phase transitions.
For the case of ferromagnetism, for example, we can transform the matrix Green's function in spin space
\begin{align}
\hat{G}(\vec{k})\equiv \begin{bmatrix} G_\uparrow(\vec{k}) & 0 \\ 0 & G_\downarrow(\vec{k})\end{bmatrix}
\end{align}
by using the unitary matrix
\begin{align}
\hat{{\cal U}}\equiv \frac{1}{\sqrt{2}} \begin{bmatrix} 1 & -1 \\ 1 & 1 \end{bmatrix}
\end{align}
as
\begin{align}
\hat{G}_{\cal U}\equiv\hat{{\cal U}}\,\hat{G}\,\hat{{\cal U}}^\dagger= \begin{bmatrix} G_+(\vec{k}) & G_-(\vec{k}) \\ G_-(\vec{k}) & G_+(\vec{k}) \end{bmatrix}
\end{align}
with $G_\pm(\vec{k})\equiv \frac{1}{2}[G_\uparrow(\vec{k})\pm G_\downarrow(\vec{k})]$. 
Ferromagnetism is characterized by the emergence of $G_-(\vec{k})$, which is determined in the mean-field theory
by
\begin{align}
\Sigma_- = -U\frac{T}{{\cal N}_{\rm a}}\sum_{\vec{k}}G_-(\vec{k})
\label{Sigma_-}
\end{align}
with $\Sigma_-\equiv \frac{1}{2}(\Sigma_\uparrow- \Sigma_\downarrow)$.
Using Eq.\ (\ref{deltaG}) to linearize $G_-(\vec{k})$ in terms of $\Sigma_-(\vec{k})$,
we obtain $\delta G_-(\vec{k})$ that is first order in $\Sigma_-$ as
\begin{align}
\delta G_-(\vec{k}) =G(\vec{k})G(\vec{k})\Sigma_-
\end{align}
with $G(\vec{k})\equiv G_+(\vec{k})$.
Substitution of it in Eq.\ (\ref{Sigma_-}) yields the mean-field equation to determine $T_{{\rm c}0}$ 
of ferromagnetism called the Stoner criterion\cite{Stoner38, Blundell01} as
\begin{align}
U\chi_{\rm ph}^0(\vec{0})=1,
\label{Stoner}
\end{align}
where $\chi_{\rm ph}^0$ represents the particle-hole bubble defined by
\begin{align}
\chi_{\rm ph}^0(\vec{q})\equiv -\frac{T}{{\cal N}_{\rm a}}\sum_{\vec{k}}G_0(\vec{k}+\vec{q})G_0(\vec{k}) .
\end{align}

\section{$\Phi$ functional of Haussmann {\it et al}.\label{Appendix-Haussmann}}

The $\Phi$ functional adopted by Haussmann {\it et al}.\ \cite{HRCZ07}
can be expressed in the present notation by using the (1,1) submatrix of Eq.\ (\ref{chi^(0c)}),
\begin{align}
\hat{\chi}_{\rm pp}(\vec{q}) \equiv \begin{bmatrix}
\vspace{1mm}
-\chi_{G\bar{G}}(\vec{q}) & -\chi_{FF}(\vec{q})  \\ -\chi_{FF}(\vec{q})  & -\chi_{\bar{G}G}(\vec{q}) 
\end{bmatrix} 
\label{hatChi_pp}
\end{align}
as
\begin{align}
\Phi_{\rm HRCZ}=\frac{T}{2}\sum_{\vec{q}}\ln \bigl[\hat{1}+U\hat{\chi}_{\rm pp}(\vec{q})\bigr] .
\label{Phi_HRCZ}
\end{align}
See Eqs.\ (2.16) and (2.18) of their paper.\cite{HRCZ07}
This functional $\Phi_{\rm HRCZ}$ consists of the series of the fourth power of $F$
so that the differentiation $\delta \Phi_{\rm HRCZ}/\delta F(\vec{k})$ does not yield any term linear in $F$ 
that affects the $T_{\rm c}$ equation.
It should also be noted that the fault in their formalism of not satisfying the Ward identities, which they pointed out around Eqs.\ (2.55) and (2.66)
of their paper,\cite{HRCZ07}
originates not from the LW formalism itself, contrary to their statement,
but rather from their incorporating only partial processes in terms of $F$ into $\Phi$.
In contrast, our formalism developed in Ref.\ \onlinecite{Kita11} does satisfy the Ward identities, as can be proved in the same manner
as that given for Bose systems in Ref.\ \onlinecite{Kita21-2},
owing to the fact that all the anomalous processes derivable from a given normal-state diagram 
are incorporated.

\section{Derivation of Eq.\ (\ref{T_c0^c-WC}).\label{Appendix-C}}

It follows from Eqs.\ (\ref{Tc0-eq3}) and (\ref{HS}) that the equality
\begin{align}
\chi_{\rm pp}^0(\vec{0})\bigr|_{T=T_{{\rm c}0}}=\chi_{\rm pp}(\vec{0})\bigr|_{T=T_{{\rm c}0}^{\rm c}}
\end{align}
holds with $\chi_{\rm pp}(\vec{0})\equiv -\chi_{G\bar{G}}(\vec{0})$.
Subtracting $\chi_{\rm pp}^0(\vec{0})\bigr|_{T=T_{{\rm c}0}^{\rm c}}$ from both sides
and transforming the left-hand side in the same way as deriving Eq.\ (\ref{Tc0}) from Eq.\ (\ref{Tc0-eq3}),
we obtain
\begin{align}
 \ln \frac{T_{{\rm c}0}^{\rm c}}{T_{{\rm c}0}}=\frac{1}{N(0)}\Bigl[\chi_{\rm pp}(\vec{0})-\chi_{\rm pp}^0(\vec{0})\Bigr]_{T=T_{{\rm c}0}^{\rm c}} ,
 \label{T_c0^c-eq}
\end{align}
The right-hand side is a function of $U$ and will approach zero quadratically in $U$;
the term linear in $U$ is canceled by the renormalization of $\mu$.
On the other hand, it follows from Eq.\ (\ref{Tc0}) that $U$ can be expressed in terms of $T_{{\rm c}0}$ as
\begin{align}
U=-\frac{\varepsilon_{\rm F}^0/N(0)}{\ln (C\varepsilon_{\rm F}^0/T_{{\rm c}0})}.
\end{align}
Now, let us approximate the left-hand side of Eq.\ (\ref{T_c0^c-eq}) in the weak coupling as $\ln \frac{T_{{\rm c}0}^{\rm c}}{T_{{\rm c}0}}\approx \frac{T_{{\rm c}0}^{\rm c}-T_{{\rm c}0}}{T_{{\rm c}0}}$ and also expand the right-hand side in $U$. We thereby obtain Eq.\ (\ref{T_c0^c-WC}).

\end{document}